# Controlled Experiments with Student Participants in Software Engineering
## Preliminary Results from a Systematic Mapping Study

Marian Daun, Carolin Hübscher, Thorsten Weyer
paluno – The Ruhr Institute for Software Technology
University of Duisburg-Essen
{marian.daun, carolin.huebscher, thorsten.weyer}@paluno.uni-due.de

## ABSTRACT
**[Context]** In software engineering research, emphasis is given to sound evaluations of new approaches. While industry surveys or industrial case studies are preferred to evaluate industrial applicability, controlled experiments with student participants are commonly used to determine measurements such as effectiveness and efficiency of a proposed approach. **[Objectives]** In this paper, we elaborate on the current state of the art of controlled experiments using student participants. As student participants are commonly only reluctantly accepted in scientific communities and threats regarding the generalizability are quite obvious, we want to determine how widespread controlled experiments with student participants are and in which settings they are used. **[Methods]** This paper reports on a systematic mapping study using high-quality journals and conferences from the software engineering field as data sources. We scanned all papers published between 2010 and 2014 and investigated all papers reporting student experiments in detail. **[Results]** From 2788 papers under investigation 175 report results from controlled experiments. 109 (62.29%) of these controlled experiments have been conducted with student participants. Most experiments used undergraduate student participants, recruited students on a voluntary basis, and set them tasks to measure their comprehension. However, many experiments lack information regarding the students' recruitment and other important factors. **[Conclusions]** In conclusion, student participation in software engineering experiments can be seen as a common evaluation approach. In contrast, there seems to be little knowledge about the threats to validity in student experiments, as major drivers such as the recruitment are not reported at all.

## 1. Introduction

For a holistic evaluation of proposed approaches in software engineering research, a combination of several empirical methods is needed to provide sound evidence regarding the value of the approach. Hence, the commonly preferred industrial research (i.e. industry surveys or industrial case studies) needs to be combined with high-quality experiments (cf. [1]). Usually, experiments with industry participants are not feasible, as typically an insufficient number of practitioners are available (cf. [2]). Furthermore, experiments in industry often lack the needed control regarding measurements and instrumentation (cf. [3]). Hence, controlled experiments with student participants are frequently used in software engineering. In addition, researchers argue that only in experiments with student participants both approaches under investigation (i.e., the proposed new approach and the common approach) are treated fairly (cf. [2]) because in industrial research, participants will usually be more trained with the actually used approach.

Research has been done on the question whether student participants are appropriate experiment subjects. In particular, it is broadly discussed whether student participation is applicable to an industrial setting (e.g., [4], [5], [6], [7]). To aid discussions regarding student participation in controlled experiments in the software engineering field, this paper contributes results of a mapping study to assess the current state of the art regarding controlled experiments with student participants. To achieve this goal, this paper reports on major aspects of student participation. One major aspect is the current state of use, i.e., it is important to provide data to investigate how common experiments with student participants are and what the typical experimental tasks and setups are. Another major aspect aims at the reporting of threats to validity. While in the past shortcomings have been identified it is to question whether the situation has improved and particularly whether researchers are aware of threats specific for student use such as the recruitment strategy. Furthermore, we investigate whether experiments with graduate participants and with undergraduate participants differ, as undergraduate experiments are commonly seen as inappropriate (e.g., [6]).

The literature review presented in this paper basically repeats the review reported in [8] for a current time frame and focusing on student participants. In particular, publications in selected high-quality journals and conferences of the years 2010 to 2014 (including) are reviewed. In total, the mapping study is based on the data provided by 109 papers which report 163 controlled experiments with student participants.

The remainder of the paper is structured according to guidelines for reporting literature reviews and mapping studies, as outlined in [9] and [10]. Hence, Section 2 discusses related reviews and research, Section 3 defines the research questions, and Section 4 details the review method and search strategy. Section 5 presents the results of the literature review, which are discussed in Section 6. To this end, Section 7 concludes the paper.

## 2. Background and Motivation

In literature, work has been done to define best practices for conducting experiments in software engineering (e.g., [11]). Thereby, general approaches also discuss issues w.r.t. student participants (i.e., how to conduct a proper experiment in an educational setting, e.g., [12], [8], [13], [14]). Furthermore, general research regarding avoidance of threats to validity has been done in the field (e.g., [15], [16], [17], [18]). Another broadly discussed aspect of experimentation in software engineering focuses on the use of student participants under the research of question whether student participants are generalizable to professionals in an industrial setting (e.g., [4], [5], [6], [7]).

Apart from work regarding best practices and detailed questions on how to conduct experiments in software engineering, work has been done aiming at state of the art analysis w.r.t. experimentation in software engineering research. *Lukowicz et al.* report in [19] a quantitative study to evaluate the state of the art of experimental evaluation in computer science for the years 1991-1994, in which a total of 403 research papers was investigated. In conclusion, about 40% of the investigated papers lacked empirically validated results although the authors argued the proposed approaches were in need of empirical evaluation. In [20], *Zelkowitz & Wallace* compared published papers in the software engineering field for the years 1985, 1990, and 1995. In total, 612 papers were under investigation. The data show that the amount of papers without experimentation was declining from 36.4%, to 29.2% and finally to 19.4%. Also *Zendler* [21] and *Glass et al.* [22] conducted a state of the art analysis of software engineering literature in 2001 and 2002. In both cases the research agenda was much broader and the use of experimentation in software engineering was just one aspect of the investigation.

Literature reviews regarding the state of the art for experiments in software engineering were reported by *Sjøberg et al.* in 2005 and *Kampenes et al.* in 2009. In [8], *Sjøberg et al.* report on a literature survey of controlled experiments in software engineering. Among 5,453 scientific publications, 103 papers reported on controlled experiments. These controlled experiment reports were analyzed in more detail to elicit the state of the art in controlled experiments as of 2005. In doing so, among others, the discussion of threats to validity and the recruitment strategies were investigated. The authors concluded, that a large amount of experiment reports lack a proper discussion of threats to validity and do not report influential factors in the designs of the controlled experiments. *Kampenes et al.* conducted a systematic review of quasi experiments in software engineering [23]. The analysis builds upon the identified controlled experiments by [8]. Hence, no newer experiment reports were considered.

## 3. Research Questions

The goal of this mapping study is to assess the current state of the art in controlled experiments with student participants in the field of software engineering. To do so, we defined three major research questions, which will be detailed by sub-questions in the respective subsections.

First, it is of importance to assess the current use of controlled experiments with student participants:

*RQ1: What is the current state of use of controlled experiments with student participants?*

Second, *Sjøberg et al.* [8] reported that many experiment reports do not report necessary aspects of the threats to validity in general, and even that the situation is worse for experiments conducted with student participants. Hence, it is to question how the situation evolved:

*RQ2: What is the current state for reporting threats to validity in controlled experiments with student participants?*

Last, experimental research with undergraduate participants is commonly seen as inadequate for software engineering research, due to a lack in generalizability. Hence, it is to question, how experiments with undergraduate and graduate participants differ.

*RQ3: What is the difference in the current state of use for controlled experiments with undergraduate and with graduate participants?*

### 3.1 Current State of Use (RQ1)

Generalizability is commonly considered a major threat to experiments with student participants. Therefore, experiments with student participants are often only reluctantly accepted in scientific communities (cf. [2]). However, due to the lack of sufficient participating practitioners they are often the only way to gather quantitative data (cf. [3]), as industry surveys and case studies often cannot be used to provide evidence for e.g., effectiveness of a proposed technique compared to an existing technique. In addition, it has recently been discussed whether the evaluation of new techniques in an industrial setting is fair (e.g., [2], [24]), since industry professionals commonly have extensive knowledge and experience with existing techniques but tend to be unfamiliar with the newly proposed technique (cf. [2]). Additionally, new approaches (such as the model-based engineering paradigm) are not incorporated into industrial practice everywhere at the same time but are introduced for specific teams successively. These teams can commonly be considered consisting of rather new employees who are in many ways comparable to graduate students (cf. [2]). Hence, it is of interest to determine to which degree controlled experiments with student participants are used:

*RQ1.1: Are controlled experiments with student participants well established in literature and commonly reported?*

*RQ1.2: What is the share compared to experiments with industry professionals?*

Regarding the current use of student experiment, it is of course also to questions what student participants are used for. Hence, it is to question whether there are some experimental tasks commonly given in experiments with student participants. This could indicate some areas where student participants are appropriate.

*RQ1.3: What are categories of experimental tasks commonly given in experiments with student participants?*

A further point in our investigation deals with the time consumption/the time given for the completion of a task. In particular, for student participants, their attention spans have to be taken into account (cf. [25]). Long lasting tasks will lead to a higher abortion rate or, in mandatory experiments, to a higher *mental* abortion rate. This means that students still complete their tasks but after a certain point in time are so uninterested that results are no longer reliable. This leads to:

*RQ1.4: What time is consumed for task completion in student experiments on average?*

### 3.2 Current State of Threats to Validity Reporting (RQ2)

As *Sjøberg et al.* [8] reported that experiment reports fail to report threats to validity in general and for student experiments in particular, it is to question whether this situation has since. Hence, it is of interest whether the threats to validity are properly discussed within the experiment reports. Considering the large number of possible threats (cf., e.g., [15], [17], [18]), we focus on the broader discussion of general validity categories. Threats to validity relate to four different categories: internal, external, construct, and conclusion validity (cf. [15]). Hence, it is to question whether the authors discuss threats to validity on all categories or whether they deem some particular validity aspect to be more important in their experiment.

*RQ2.1: What categories of threats to validity are explicitly discussed in the experiment reports?*

In controlled experiments with student participants, the recruitment is directly linked to several threats to validity (cf. [13], [12]). For example, given the use of bonuses such as grade improvement, can lead to increased performance in experiments compared to the typical performance under normal conditions (cf. [26], [27]). Furthermore, this may lead to threats from hypothesis guessing as students are eager to answer as desired by the researcher (cf. [28]). In contrast, mandatory participation is often seen as ethically abusive (cf. [29]), unless the experiment is carefully designed to aid the students' learning experience of course related topics (e.g., [30], [31], [32]). Hence, it is of interest whether these points are known (i.e., is the recruiting discussed in experiment reports) and which recruiting strategies are commonly chosen:

*RQ2.2: Is the recruitment reported in experiments with student participants?*

*RQ2.3: What are commonly used recruiting strategies?*

A specific part of the recruitment strategy deals with bonuses given for participation. While monetary payment is often seen as the best way to reward students (cf. [13]), according to the literature, many experiments make use of grading improvement, which entails the risk of misinterpreting student results (cf. [12]).

*RQ2.4: Are the bonuses given reported in experiments with student participants?*

*RQ2.5: Which kind of bonuses are given for experiment participation?*

The choice of the experimental control (e.g., within subject, control groups) has major effects on the threats to validity of an experiment and on the participants' results (cf. [33]). In the case of student experiments, the use of control groups can be infeasible, as in teaching environments it needs to be ensured that all participants get the possibility to gain the same knowledge in experiments (cf. [29]). Hence, it is to question whether specific setups are commonly chosen in experiments with student participants.

*RQ2.6: What are commonly chosen control strategies in the experimental setups of experiments with student participants?*

## 3.3 Difference in Current State of Use for Graduates and Undergraduates (RQ3)

In literature, a differentiation is commonly made between student experiments with undergraduate participants and graduate participants. With respect to generalizability, it is often stated that graduate students are appropriate study participants in some cases (e.g., [4], [5], [6]) or even most of the time (cf. [7]). In contrast, undergraduate students seem to be regarded as inappropriate in general (e.g., [34], [6]). Therefore, it is of interest whether these statements are accordingly reflected in literature:

*RQ3.1: Are controlled experiments with undergraduate student participants uncommon?*

*RQ3.2: What is the ratio between student experiments with graduate participants and with undergraduate participants?*

Furthermore, findings indicate that the participants' degree of experience and ability has major effects on the task completion in experiments (cf. [35], [36]).

*RQ3.3: Do the tasks completed by students in controlled experiments differ for experiments with undergraduate students and for experiments with graduate students?*

For the obvious reason of completeness, it is also to question whether the reporting of threats to validity differs:

*RQ3.4: Does the reporting of threats to validity, recruitment, bonuses, and control differ for experiments with graduate students and for experiments with undergraduate students?*

## 4. Review Methods

### 4.1 Data Source and Search Strategy

In our study, we investigated controlled experiments reported in a set of high-quality venues. In particular, we adopted the search strategy used by *Sjøberg et al.* [8] to investigate the state of the art w.r.t. the reporting of controlled experiments. Hence, we chose our data sources close to [8]. Differences in the data sources result from limiting factors regarding the quality assessment of the publications under investigation. Detailed information is provided in Section 4.3.

In total, we reviewed publications in the following journals:

- ACM Transactions on Software Engineering Methodology (TOSEM),
- Empirical Software Engineering (EMSE),
- IEEE Transactions on Software Engineering (TSE),
- Information and Software Technology (IST), and
- Journal of Systems and Software (JSS).

In addition, we reviewed publications in the following conferences:

- Empirical Assessment & Evaluation in Software Engineering (EASE), and
- International Symposium on Empirical Software Engineering and Measurement.

To ensure an actual investigation of the current state of the art regarding the reporting of controlled experiments with student participants, we restricted the review to journal volumes and conference proceedings that have been published within the last five years.

### 4.2 Search Process and Experiment Selection

We investigated the discussed data sources for publications including reports of controlled experiments conducted with student participants. In order to determine the relevant experiment reports, the selection process was tripartite:

- First, all relevant volumes and proceedings of the journals and conferences under investigation were obtained. Relevant for the literature search were the volumes and proceedings published between 2010 and 2014. All relevant issues could be received from different digital libraries (i.e., IEEE *Xplore* Digital Library, ACM digital library, Springer Link). Hence, no venues had to be excluded due to non-availability.
- Second, all papers contained in the obtained volumes and proceedings were differentiated into potential relevant pa-

pers or irrelevant papers. At this point, title, keywords, and abstract were reviewed. If it became obvious that a paper was not relevant for the study, the paper was excluded from further investigation; otherwise the paper was considered a potentially relevant paper. Irrelevant papers are those papers that do not report evaluation results at all, where it is clearly stated that another investigation method than controlled experiments was used, or where obviously non-student participants were used.

- Third, the remaining potentially relevant papers were investigated in detail. In doing so, irrelevant papers were excluded and for relevant paper an analysis criteria table was filled out (see Section 4.4). A paper was excluded based on the same criteria as in the second step (i.e., the paper did either not report a controlled experiment or the experiment reported did not use student participants).

### 4.3 Experiment Quality Assessment

As briefly outlined in Section 4.1, we excluded some data sources used by [8] from our investigation. This includes journals such as IEEE Computer and IEEE Software as well as conferences with a technical focus on specific software engineering areas. In these cases, experiment reports were typically so briefly summarized that it was neither possible to determine whether our main criteria (e.g., controlled experiment, student participants) are fulfilled, nor was it possible to extract data for answering our review questions as commonly insufficient data was provided by experiment summaries. Hence, no papers of these venues were part of the investigation.

### 4.4 Inclusion and Exclusion Criteria

To ensure qualitative statements regarding the review questions for the current time only papers were **included**, which

- were published in the data sources of Section 4.1,
- were published within the last five complete years (i.e., in the years 2010, 2011, 2012, 2013, and 2014),
- are experiment reports or included an evaluation section describing a controlled experiment, and
- are clearly using students as experiment participants.

As already introduced we **excluded** entire venues from the evaluation where commonly too little information was provided regarding the review questions. Namely, there were pieces of information missing regarding

- the recruitment of student participants,
- the detailed tasks students were asked to complete,
- the time frame given for experiment completion,
- the time frame intendedly needed for task fulfillment,
- the experimental design, or
- the threats to validity of the experiment.

Note that also papers of the chosen data sources did not report one or multiple (in some cases even all) previously mentioned information. In these cases, we did not exclude the paper from our investigation but recorded that the particular aspect is unknown.

### 4.5 Data extraction

The selected relevant papers were read and statements regarding certain criteria (e.g., the recruitment, bonuses given, graduate or undergraduate participants) were excerpted. To do so, a table containing all relevant experiments (and their reporting papers) was filled out. Each cell was filled with the information provided as long as some information was given. For example, in case the recruitment strategy was neither reported explicitly in the participants Section or the threats to validity Section nor reported implicitly anywhere else throughout the paper, the cell remained empty.

Subsequently, the table was reviewed and the documented information was categorized by different researchers. In case categories were distinct, the matter was discussed and collectively solved. Table 1 shows criteria and final categories. As can be seen, in some cases categories are exclusive (e.g., students could either be recruited voluntarily or mandatorily in one experiment) while in other cases categories are non-exclusive (e.g., as experimental participants, undergraduate students as well as graduate students could be used in the same experiment). Some categories also depend on other categories, for example, an experiment can only use professional participants in case also students are used as participants, otherwise the experiment would be excluded due to our exclusion criteria (see Section 4.4).

**Table 1. Experiment categorization**

| Category | Values |
| --- | --- |
| Type of participants (non-exclusive) | Graduates |
| | Undergraduates |
| | Professionals (only in mixed Experiments) |
| Number of Participants | INTEGER |
| Recruitment (exclusive) | Voluntary |
| | Mandatory |
| | Unknown |
| Bonuses and Rewards (exclusive) | Money (i.e. paid participants) |
| | Bonus to the Exam or the Final Points |
| | No Bonuses |
| | Unknown |
| Task Type (non-exclusive; open) | Comprehension |
| | Evaluation |
| | Maintenance |
| | Testing |
| | Programming |
| | Modelling |
| Task Duration | TIME |
| Experiment Control (exclusive) | Control Group(s) |
| | Within-Subject Design |
| | No Control |
| | Unknown |
| Threats to Validity (non-exclusive) | Internal Validity |
| | External Validity |
| | Conclusion Validity |
| | Construct Validity |

# 5. Results

This Section provides quantitative data regarding the research questions. Each subsection relates to a research questions.

In total, we investigated 2788 papers. 109 of these papers under investigation have been included according to our inclusion criteria from Section 4.3. These 109 papers describe 163 controlled experiments, as, for example, a controlled experiment and its replication, which is also a controlled experiment, are reported in the same paper.

## 5.1 Current State of Use (RQ1)

### 5.1.1 Share of Student Experiments (RQ1.1, RQ1.2)

Table 2 shows the distribution of publications, experiment reports in general, and experiment reports with student participants in specific. As can be seen, the reporting of controlled experiments in general can be seen as common for most venues Furthermore, the use of students as participants in controlled experiments in the software engineering field can be described as very common.

In comparison, over the years of investigation, an increasing number of papers reporting controlled experiments with student participants is recognizable. While in 2010 only 13 controlled experiments with student participants have been reported in journals and conferences under investigation, in 2014, controlled experiments with student participants were reported in 33 papers. First impressions of the 2015 editions of the journals and conferences indicate that this trend is continuing.

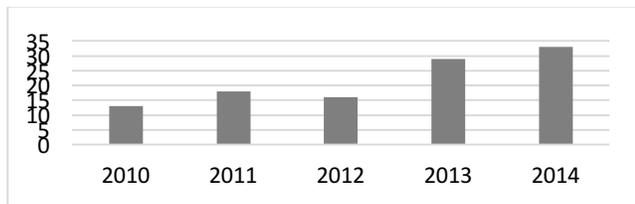

**Figure 1. Annual distribution of papers reporting experiment reports with student participants**

### 5.1.2 Tasks in Student Experiments (RQ1.3)

We classified the tasks completed by students in the reported experiments. The majority of tasks deal with the students' comprehension (e.g., asking questions regarding the understanding of a model or program code). Evaluation tasks (e.g., reviews and inspections), maintenance tasks (e.g., incorporating new requirements into models or code), and testing tasks (e.g., detecting errors in a prototype) are also commonly used in student experiments (see Figure 2). In detail, 35.58% of reported experiments ask to complete comprehension tasks, 20.85% ask for testing tasks, 20.25% ask for maintenance tasks, and 15.95% ask for evaluation tasks. Further 15.34% of reported experiments ask to complete tasks which can be considered highly individual as no other controlled experiment contains a similar task.

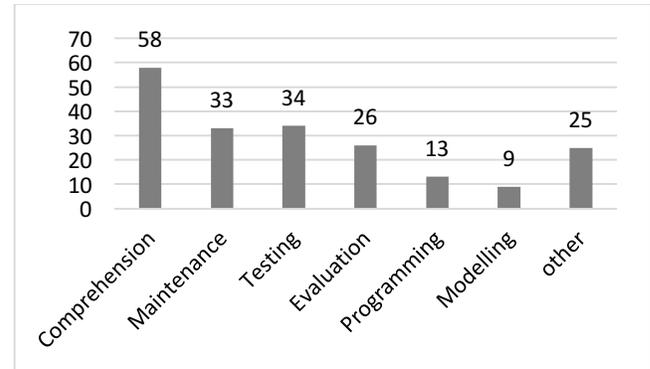

**Figure 2. Tasks to be completed in student experiments**

### 5.1.3 Time Consumption for Tasks (RQ1.4)

Figure 3 shows the average time consumption given for task completion. As can be seen, the majority of experiments is designed to be completed in more than two hours (30.06%) or even in unlimited time (33.13%). Only 1.84% are designed to be completed in less than thirty minutes and 20.86% are designed to be completed in less than two hours. 14.11% of experiment reports do not give information regarding the time consumption for task completion.

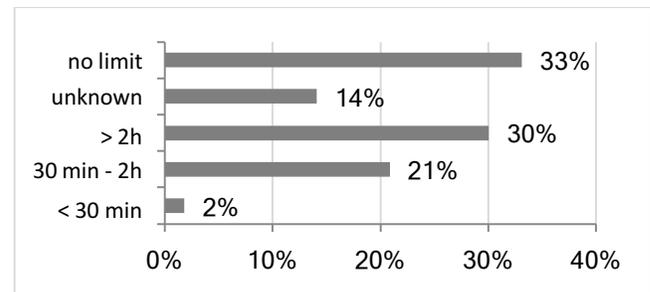

**Figure 3. Given time for task completion**

**Table 2. Distribution of controlled experiments and controlled experiments with student participants in the respective venues**

|  | publications in venue | | reports of experiments with human participants | | reports of experiments with student participants | |
|---|---|---|---|---|---|---|
|  | absolute | % of all venues | absolute | % in venue | Absolute | % of experiment reports |
| TOSEM | 136 | 4.88% | 15 | 11.03% | 11 | 73.33% |
| EMSE | 250 | 8.97% | 48 | 19.2% | 26 | 54.17% |
| TSE | 348 | 12.48% | 19 | 5.46% | 11 | 57.89% |
| IST | 578 | 20.73% | 28 | 4.84% | 18 | 64.29% |
| JSS | 1016 | 36.44% | 23 | 2.26% | 12 | 52.17% |
| EASE | 163 | 5.85% | 16 | 9.82% | 12 | 75.0% |
| ESEM | 297 | 10.65% | 26 | 8.75% | 19 | 73.08% |
| total | 2557 | 100 % | 175 | 6.28% | 109 | 62.29% |

## 5.2 Current State of Threats to Validity Reporting (RQ2)

### 5.2.1 Reported Threat to Validity Categories (RQ2.1)

The experiment reports under investigation most commonly discuss threats to external and internal validity. Threats to both validity categories have been reported in the vast majority of experiment reports (i.e., external validity was discussed in 90.74% of experiment reports and internal validity was discussed in 90.12% of experiment reports). Threats to construct validity were discussed in 73.46% of all experiment reports and threats to conclusion validity were discussed in 64.2% of all experiment reports. It must be noted that 7.41% of all reported experiments with student participants did not discuss threats to validity at all.

Figure 4 shows the threats to validity discussed in experiment reports with respect to common categories for threats to validity. It is to note that some authors did not refer to the categories as suggested by [15] but only differentiate between threats to external validity and threats to internal validity. Therefore, we classified the reported and discussed threats to validity from the experiment reports according to [15] if authors used a different categorization.

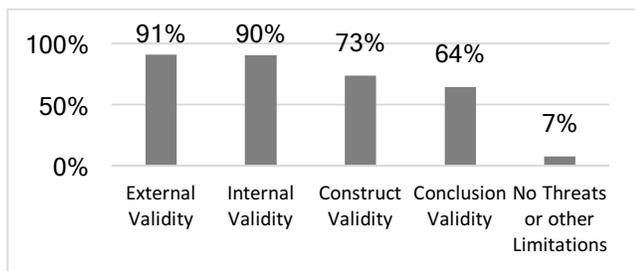

**Figure 4. Threats to validity**

### 5.2.2 Recruitment Strategy (RQ2.2, RQ2.3)

Regarding student recruitment, in 46.63% of experiments with student participants, students were recruited voluntarily, in about 15.95% students were recruited on a mandatory basis, and 37.42% of experiment reports did not report the recruitment strategy (see Figure 5). It must be noted that in the case of mandatory recruitment, two experiments reported in distinct papers are also included where students were not aware of their experiment participation. In these cases, the experiment was part of the graded exam but students were unaware that they participated in an experiment.

While the number of papers not reporting the recruitment is increasing over the years (i.e., 3 in 2010 and 26 in 2014), the ratio between controlled experiments reporting and not reporting the recruitment can be considered rather unchanging.

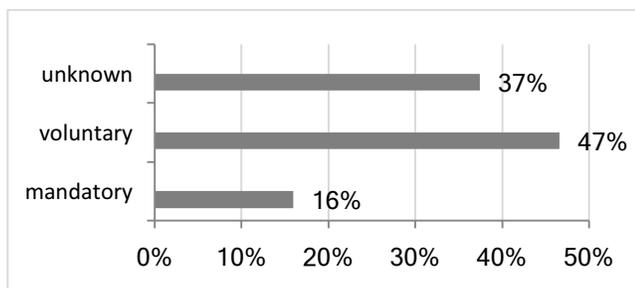

**Figure 5. Student recruitment**

### 5.2.3 Bonus System (RQ2.4, RQ2.5)

It is often common to provide bonuses to ensure student participation and furthermore gain student commitment for task completion. In the majority of the experiment reports, no statements regarding bonuses are made (54.60%), but this might correspond with researchers not stating what they have not done. For example, researchers might assume that mandatory recruitment also means that no bonuses are given. Overall, only 15.95% of experiments reported did explicitly refrain from rewards. 23.31% explicitly used bonuses related to grading, which is commonly seen as a severe threat to the internal validity of the experiments (cf. [12]). Regarding the development over the years, also in the case of rewards given no clear trend is recognizable.

Figure 6 details the distribution of the kind of bonuses given. As can be seen, only a small minority makes use of monetary payment (in total, 1.84% or 3 experiments). Since this bonus system is often appreciated w.r.t. ensuring voluntary recruitment and minimizing threats resulting from favoring the researcher (cf. [13]), it is interesting to note, that this bonus system has hardly any relevance in practice. The category "other" relates to minor gifts such as sweets or participation in some kind of lottery. In some cases other benefits are combined with grade improvements. In these cases the experiments were categorized as "grades or extra points".

Figure 7 details the relation between the recruitment strategy and the rewards given. As can be seen, experiments without a reported recruitment strategy often also lack providing the information regarding bonuses, too. In detail, 68.85% of experiments not detailing the recruitment strategy, did also not provide information regarding the incentive system used nor did they report that no bonuses were given.

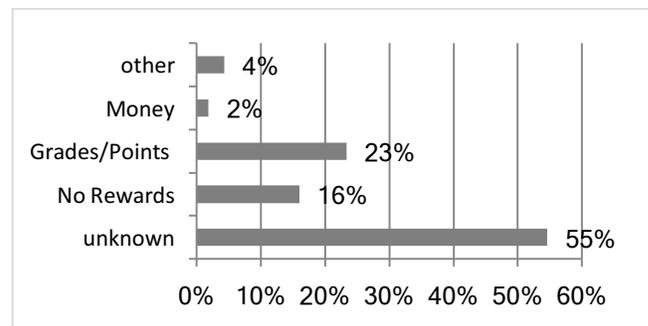

**Figure 6. Rewards given**

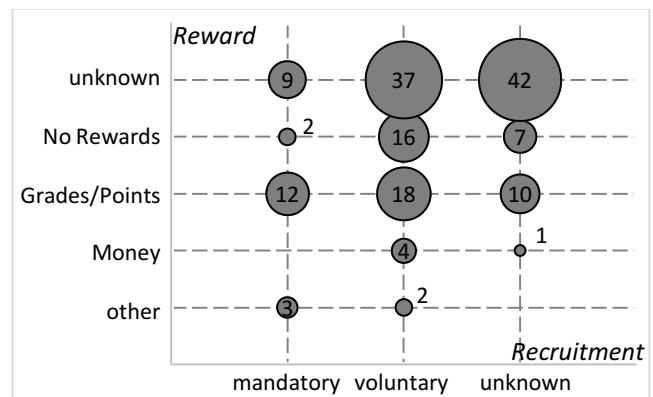

**Figure 7. Rewards and recruitment**

### 5.2.4 Control Used in Student Experiments (RQ2.6)

The vast majority (77.6%) of controlled experiments with student participants uses an experimental design which makes use of distinct treatment- and control-groups (see Figure 8). It must be positively noted that only 6.4% do not discuss whether they used a control and only 3 experiments did explicitly use no control.

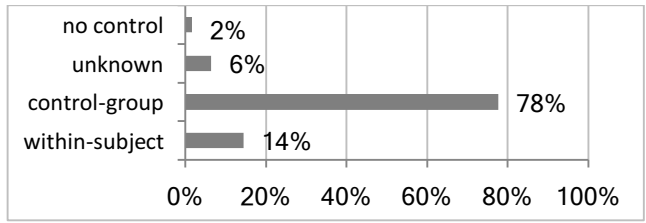

**Figure 8. Distribution of graduate and undergraduate participants in controlled experiment reports**

Only 14.4% make use of a within-subject design. This is in so far interesting, as it is often discussed that experiments in educational settings must provide the same learning experience for all participants (cf. [29]). Figure 9 compares the control used to the recruitment strategy chosen. Only one mandatory experiment uses a within-subject design but 24 use a design with a control-group.

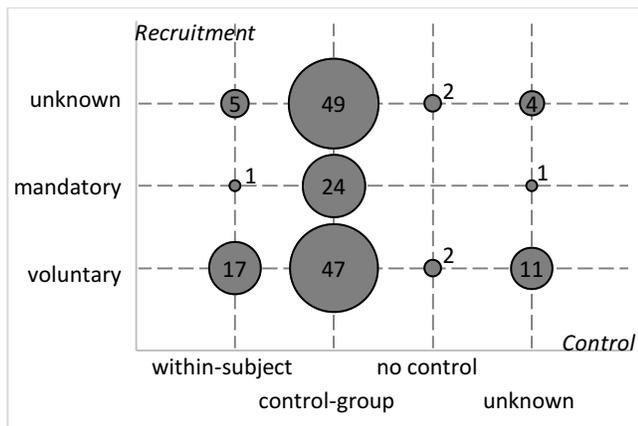

**Figure 9. Recruitment and control**

## 5.3 Difference in Current State of Use for Graduates and Undergraduates (RQ3)

In this section, a comparison between controlled experiments with graduate participants and undergraduate participants is made. Hence, mostly experiments with graduate student participants only and with undergraduate participants only are compared.

### 5.3.1 Ratio of Experiments with Graduate and Undergraduate Student Participants (RQ3.1, RQ3.2)

Most experiments with student participants explicitly stated that experiments were conducted with graduate student participants only (42.33%). In addition, 34.36% explicitly used undergraduate student participants only. 4.29% of reported student experiments did not classify their participants w.r.t. their level of maturity. 7.98% explicitly used graduate as well as undergraduate students. In addition, 11.04% of the reported experiments with student participants not only used student participants alone but a combination of student and professional participants. Figure 10 visualizes the distribution in percentage and absolute numbers.

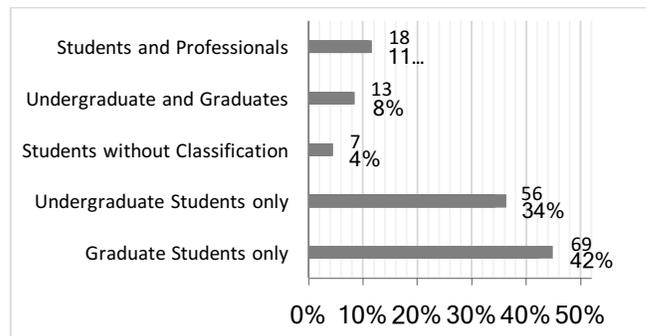

**Figure 10. Distribution of graduate and undergraduate participants in controlled experiment reports.**

Figure 11 shows how many students in average participated in an experiment. The figure illustrates the differences regarding the number of student participants for each experiment dependent on their maturity level. In mean, graduate experiments have about 23.5 participants, while undergraduate experiments have 33. These findings are not surprising, as in university education, undergraduate courses are typically far bigger than graduate courses. Participant numbers reach up to 270 in undergraduate experiments as opposed to up to 196 participants in graduate experiments.

In three cases, we could not consider the experiments in this calculation as the papers did not state the number of participants. In two papers it was discussed in the threats to validity that the number of participants has to be considered very low and one paper only stated that the number of participants was sufficient.

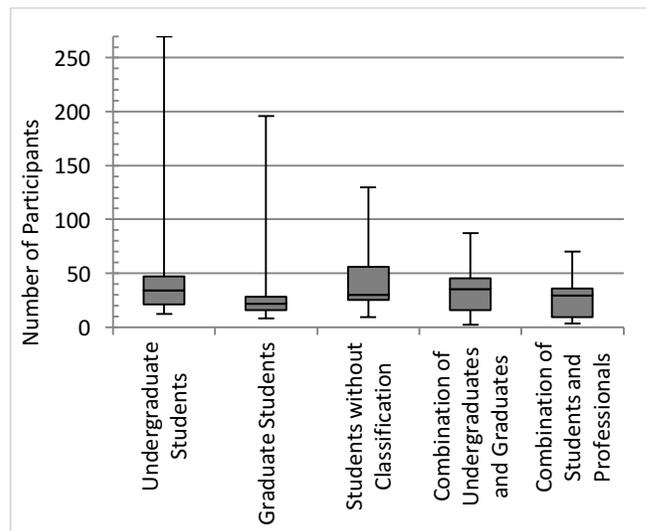

**Figure 11. Distribution of participant numbers in graduate and undergraduate experiments**

### 5.3.2 Differences in Tasks (RQ3.3)

Regarding the question of whether the tasks differ for graduate and undergraduate students, this seems to be true. While most controlled experiments with student participants aim at the students' comprehension, this is even more significant for undergraduate experiments. In contrast, graduates are more often given more specific tasks such as maintenance or evaluation tasks. Figure 12 visualizes the results for the task categories.

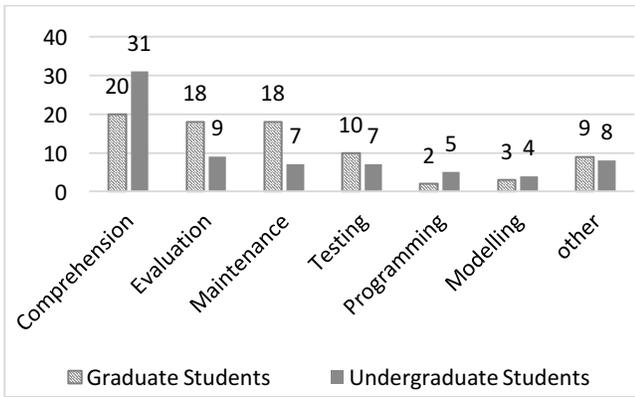

**Figure 12. Task categories in dependence of participants**

The time-frames given for the completion of experiments with graduate and undergraduate do not differ at large, as can be investigated in Figure 13.

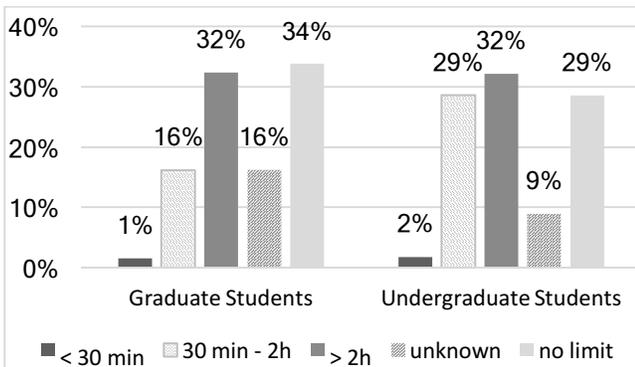

**Figure 13. Time given in dependence of participants**

### 5.3.3 Differences in Threats to validity (RQ3.4)

There is no significant difference in reporting threats to validity for controlled experiments conducted with graduate students as participants and with experiments conducted with undergraduate students as participants. By tendency, controlled experiments with undergraduate participants do more often report threats to validity in each category. Figure 14 shows the reported threat to validity categories depending on the participating students.

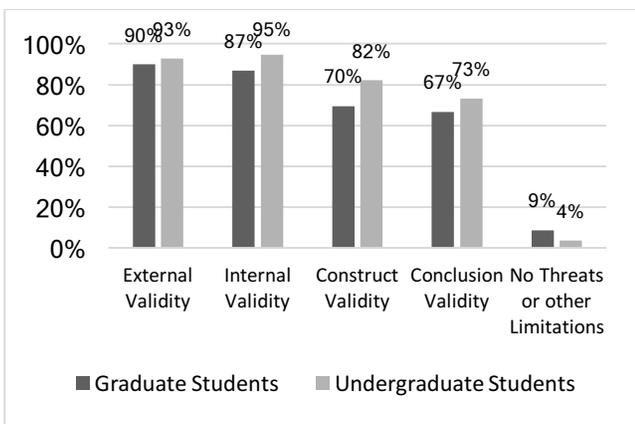

**Figure 14. Reported threats to validity categories in dependence of participants**

The ratio between voluntary and mandatory recruitment differs between controlled experiments with undergraduate and graduate participants. Experiments with graduate student participants do more often not report the recruitment strategy use less often voluntary recruitment compared to experiments with undergraduate student participants. The differences are shown in Figure 15.

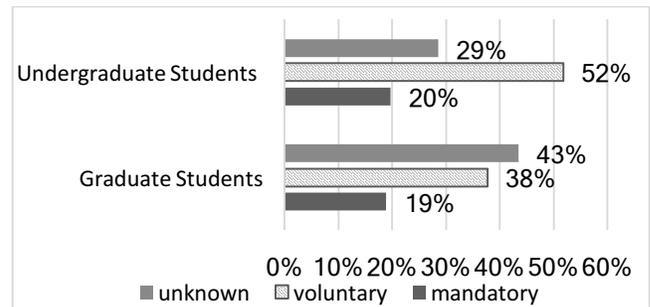

**Figure 15. Recruitment in dependence of participants**

Regarding the rewards given, experiments with undergraduate participants only make more often use of grade improvements in comparison with controlled experiments with graduate participants (37.50% vs 20.29%, see Figure 16). Neither experiments with graduates only nor experiments with undergraduates only make use of monetary payment as reward for participation.

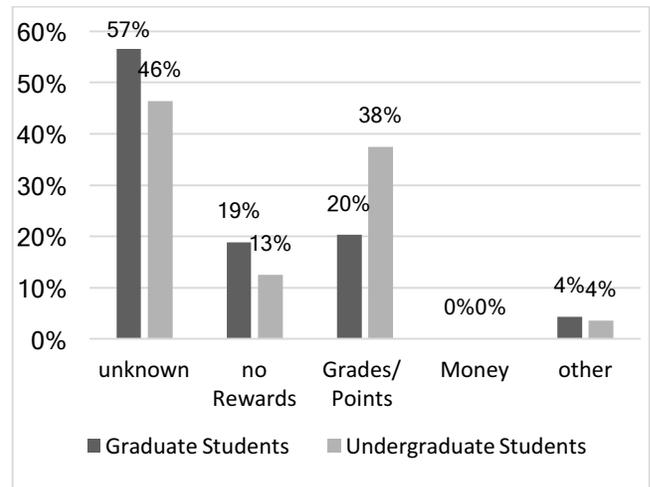

**Figure 16. Rewards in dependence of participants**

Regarding the use of control-groups, in controlled experiments with graduate students a within-subject design is more often used than in controlled experiments with undergraduate students (see Figure 17).

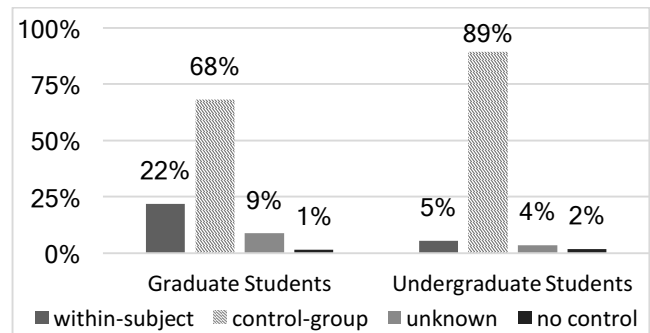

**Figure 17. Control used in dependence of participants**

# 6. DISCUSSION
## 6.1 Principal Findings
Table 3 summarizes the findings detailed in Section 5 with respect to the review questions from Section 3. In summary, we can conclude for the major research questions to assess the state of the art in controlled experiments with student participants that:

- Student participants are used in the majority of controlled experiments in the software engineering field. Hence, the use of student participants must be considered state of the art. Experiments with student participants are commonly used for comprehension tasks and in the majority of cases last longer than two hours or have no time limit. **(RQ1)**
- The reporting of threats to validity seems to have improved since the investigation of Sjøberg et al. [8]. However, the reporting of student specific factors threatening the validity of the experiment are often not reported. **(RQ2)**
- It can be recognized that undergraduates are commonly used as experiment participants and must hence be seen as appropriate participants. While there are some minor differences between experiments with undergraduate and graduate participants, experiments with undergraduates tend to be more general and less specific. **(RQ3)**

## 6.2 Strengths and Weaknesses
While we investigated the reporting of threats to validity in the controlled experiments under investigation, this section covers the threats to validity of our own investigation. Subsequently, we will discuss the major threats of the mapping study.

First of all, we must note the limited data sources. The mapping study relies on experiment reports from only a few journals and conference proceedings. As these focus on software engineering at large or empirical software engineering in particular, results may significantly differ for more topic-specific fields of software engineering. For example, a more technical field like programming or a more practice-oriented field like requirements engineering may give different importance to particular threats to validity or to the applicability of student participants.

Table 3. Principal findings for each review question

| Identifier | | Review question | Principal findings |
|---|---|---|---|
| **RQ 1** | RQ1.1 | *Are controlled experiments with student participants well established in literature and commonly reported?* | Yes, the majority of controlled experiments is conducted using student participants. The number of experiments with student participants is increasing from year to year. |
| | RQ1.2 | *What is the share compared to experiments with industry professionals?* | Between 50% and 75% of experiment reports in each venue are conducted with student participants. |
| | RQ1.3 | *What are categories of experimental tasks commonly given in experiments with student participants?* | The majority of experiments use tasks regarding the participants' comprehension. Other often used categories are maintenance, testing, and evaluation. |
| | RQ1.4 | *What time is consumed for task completion in student experiments on average?* | Only less than 2% last less than 30 minutes and about 21% last less than two hours. About 30% last longer than two hours and 33% explicitly have no time limit. |
| **RQ 2** | RQ2.1 | *What threat to validity categories are explicitly discussed in the experiment reports?* | 90.74% discuss external validity, 90.12% internal validity, 73.46% construct validity, 64.2% conclusion validity. |
| | RQ2.2 | *Is the recruitment reported in experiments with student participants?* | 37.42% of experiments do not report the recruitment strategy. |
| | RQ2.3 | *What are commonly used recruiting strategies?* | 46.63% use voluntary, 15.95% use mandatory recruitment. |
| | RQ2.4 | *Are the bonuses given reported in experiments with student participants?* | 54.6% of experiments give no information regarding the bonuses given. |
| | RQ2.5 | *Which kind of bonuses are given for experiment participation?* | 23.31% benefit the grading, 15.95% give no bonuses, only 1.84% are paid. |
| | RQ2.6 | *What are commonly chosen control strategies in the experimental setups of experiments with student participants?* | 77.6% use some design explicitly featuring some kind of control group, 14.4% use a within-subject design, and only 1.84% do not use a control instrument at all. 6.4% do not report the control chosen. |
| **RQ 3** | RQ3.1 | *Are controlled experiments with undergraduate student participants uncommon?* | No, considering mixed setups up to 50% of experiments use undergraduates. |
| | RQ3.2 | *What is the ratio between student experiments with graduate participants and with undergraduate participants?* | 42.33 % of experiments used graduate students only. 34.36% explicitly used undergraduates only. |
| | RQ3.3 | *Do the tasks completed by students in controlled experiments differ for experiments with undergraduate students and for experiments with graduate students?* | Yes, undergraduate students are more often used for comprehension tasks and less often for more specific tasks. Time consumption does not differ significantly. |
| | RQ3.4 | *Does the reporting of threats to validity, recruitment, bonuses, and control differ for experiments with graduate students and for experiments with undergraduate students?* | Yes. Regarding the reporting of threats to validity only minor differences can be recognized but for recruitment experiments with graduate students do more often not discuss the recruitment and make less often use of voluntary recruitment. Bonuses to the grades are more often given in undergraduate experiments (38% vs 20%) and experiments with graduate students use more often a within-subject design (22% vs 5%). |

Another issue deals with the classification of the experiment reports. In classifying the experiments w.r.t. the classification scheme (Section 4.5), we did not consider different levels of fulfillment. Hence, there is a risk we placed an experiment in categories erroneously due to lacking information in the paper. For example, regarding the potential bonuses given, many papers state that they did not pay the students. In these cases, it is not decidable whether this means that no bonuses were given or whether bonuses were given (which the authors did not mention) but the bonuses were not monetary. In this example, we decided to assume that no bonuses were given, as the authors were at least aware of the necessity to report on some information regarding bonuses. Otherwise, it would be necessary to distinguish between different degrees of full information given and no information given. In fact, many papers do not report all relevant facets. Consequently, a large amount of experiments had to be classified as 'information unknown' even when the authors discussed some aspects.

Another issue deals with our inclusion and exclusion criteria: the definition of a student participant. During the investigation of the selected relevant experiment reports it became obvious that some researchers classified student participants as professionals in case they are part-time employed. In some cases, this holds explicitly for dual study programs (i.e., degree programs consisting of a university taught part and industrial apprenticeship). In other cases, it is impossible to determine the degree of professional experience of student professionals. Hence, we assume that a considerable number of experiments reporting student and professional participants must be rather considered as student participants only, as the professionals in the experiments will most likely not possess sufficient experience to demarcate from students. Furthermore, it must be considered a high risk that experiments were erroneously excluded as the reported professional participants might also be students and hence not differ from experiments with student participants.

Finally, as for most mapping studies, we investigated abstract topics, such which categories for threats to validity have been reported. Hence, this paper cannot state whether the reporting is correct, whether the threats are complete, and other detailed content-related questions.

### 6.3 Meaning of Findings

Under consideration of the threats to validity we can conclude that experiments with student participants are at least common in software engineering at general. While we investigated qualitative journals and conferences, we assume that student participants will be even more often adequate in other publications such as workshop proceedings. However, reports from other researchers show a high refusal of experimental research with student participants (cf. [6], [13]), particularly it is concluded that in many conferences papers are rejected due to the use of student participants (cf. [2]). These findings might explain the low number of experiment reports at all. Only about 10% of investigated publications featured a report of a controlled experiment with human participants, and as we investigated not only experiment reports but also the evaluation section of research papers, we consider this number as low in view of the high appreciation of experiments in literature (e.g., [1], [3], [15]). Hence, a less reluctant stance against student participants might increase the number of papers using controlled experiments as an evaluation approach.

Furthermore, the ratio of reported controlled experiments with undergraduate participants shows, in contrast to common claims in literature (cf. [6], [34]), that also undergraduate experiment participants can be appropriate. This particularly holds for simple experimental tasks regarding the participants' comprehension. Considering that the vast majority of controlled experiments reported used student participants, the use of undergraduate participants must somehow be considered as the current state of the art.

Regarding the reporting of threats to validity of controlled experiments, at a first glance, the situation has improved since the investigation of *Sjøberg et al.* [8]. In contrast, the reporting of contributing factors such as the recruitment strategy or the bonus systems shows that there remains a lack of information provided by experiment reports, which has not improved since [8]. However, we assume the fact that experiment reports discuss threats to validity – and this in most cases holds for all threat categories – as improvement. This shows that the awareness of the need for reporting threats to validity has improved.

Regarding the threats specific to student participants, we assume a lack of awareness for threats aside from the generalizability of the experiment results. As the recruitment strategy has severe impact on the threats to validity of the study, it is alarming that 32.5% of experiments did not report how students were recruited. Nevertheless, this result is not surprising as *Sjøberg et al.* [8] already reported in 2005 on the lack of information regarding the recruitment process in student experiments. The same holds for the lack in reporting the bonus system.

## 7. CONCLUSIONS

In this paper, we reported a mapping study to assess the state of the art of controlled experiments with student participants in the field of requirements engineering. To do so, we reviewed 1024 papers published in selected journals and conferences between 2010 and 2014 (including). In total, we found 95 papers which report 125 controlled experiments with student participants. Thus, this is the majority of all controlled experiments published.

Findings indicate that controlled experiments with student participants and also with undergraduate participants must be considered state of the art for evaluation in software engineering. In conclusion, it can be assumed that student experiments are appropriate as evaluation for software engineering approaches and can particularly be used to test the participants' comprehension of techniques under evaluation.

It also turned out that while the overall reporting of threats to validity has improved, the reporting of concerns specific to experiments with student participants is lacking, hence suggesting that the implications of such contributing student specific factors are often unknown. Therefore, it must be suggested to not only focus the attention in student experimentation towards the problem of generalizability but thoroughly elaborate on other threats as well.

Future work should complement the mapping study by investigating more detailed facets of student experimentation. For instance, it is of interest to investigate whether the reported threats to validity correspond with the student specific factors of the experiment. In addition, the appropriateness of the reported experiments not only for research but in the view of teaching software engineering must be considered.

# 8. REFERENCES

[1] V. Basili, „The Experimental Paradigm in Software Engineering," in *Experimental Software Engineering Issues: Critical Assessment and Future Directives, LNCS 706,* Springer, 1993.

[2] I. Salman, A. T. Misirli und N. J. Juzgado, „Are Students Representatives of Professionals in Software Engineering Experiments?," *Proceedings of the 37th IEEE/ACM International Conference on Software Engineering, ICSE 2015, Florence, Italy, May 16-24, 2015. IEEE 2015, ISBN 978-1-4799-1934-5,* pp. 666-676, 2015.

[3] R. Wieringa, „Empirical research methods for technology validation: Scaling up to practice," *The Journal of Systems and Software,* pp. 19-31, 2014.

[4] M. Höst, B. Regness und C. Wohlin, „Using Students as Subjects - A Comparative Study of Students and Professionals in Lead Time Impact Assessment," *J Emp. Softw. Eng.,* pp. 201-214, 2000.

[5] M. Svahnberg, A. Aurum und C. Wohlin, „Using Students as Subjects – an Empirical Evaluation," *Proc. of ESEM,* pp. 288-290, 2008.

[6] P. Runeson, "Using Students as Experiment Subjects - An Analysis on Graduate and Freshmen PSP Student Data," *Proc. of EASE,* pp. 95-102, 2003.

[7] W. Tichy, „Hints for Reviewing Empirical Work in Software Engineering," *J. Emp. Softw. Eng.,* pp. 309-312, 2000.

[8] D. Sjøberg, J. Hannay, O. Hansen, V. Kampenes, A. Karahasanovic, N. Liborg und A. Rekdal, „A Survey of Controlled Experiments in Software Engineering," *IEEE Trans. Software Eng.,* pp. 733-753, 2005.

[9] B. Kitchenham, „Procedures for Performing Systematic Reviews," Keele University, Keele, Staffs, UK, 2004.

[10] K. Petersen, R. Feldt, S. Mujtaba und M. Mattsson, „Systematic Mapping Studies in Software Engineering," *Proc. of EASE,* pp. 68-77, 2008.

[11] A. J. Ko, T. D. LaToza und M. M. Burnett, „A practical guide to controlled experiments of software engineering tools with human participants," *Empirical Software Engineering 20(1),* pp. 110-141, 2015.

[12] J. Carver, L. Jaccheri, S. Morasca und F. Shull, „Issues in Using Students in Empirical Studies in Software Engineering Education," *Proc. of Software Metrics Symposium,* pp. 239-249, 2003.

[13] D. Sjøberg, B. Anda, E. Arisholm, T. Dybå, M. Jørgensen, A. Karahasanovic, E. Koren und M. Vokác, „Conducting Realistic Experiments in Software Engineering," *Proc. of Intl. Symp. on Emp. Softw. Eng.,* pp. 17-26, 2002.

[14] D. Berry und W. Tichy, „Response to 'Comments on Formal Methods Application: An Empirical Tale of Software Development'," *IEEE Trans. Softw. Eng.,* Bd. 29, Nr. 6, pp. 572-575, 2003.

[15] C. Wohlin, P. Runeson, M. Höst, M. Ohlsson, B. Regnell und A. Wesslén, Experimentation in Software Engineering. An Introduction, Boston/Dordrecht/London: Kluwer Academic Publishers, 2000.

[16] C. Robson, Real World Research, 3rd Hrsg., Hoboken: John Wiley & Sons, 2011.

[17] D. Campbell und J. Stanley, Experimental and Quasi-Experimental Designs for Research, Boston: Houghton Mifflin Company, 1963.

[18] T. Cook und D. Campbell, Quasi-Experimentation - Design and Analysis Issues for Field Settings, Houghton Mifflin Company, 1979.

[19] P. Lukowicz, E. A. Heinz, L. Prechelt und W. F. Tichy, „Experimental Evaluation in Computer Science: A Quantitative Study," *Journal of Systems and Software,* 1995.

[20] M. V. Zelkowitz und D. Wallace, „Experimental validation in software engineering," *Information and Software Technology 39(11),* 1997.

[21] A. Zendler, „A Preliminary Software Engineering Theory as Investigated by Published Experiments," *Empirical Software Engineering,* Bd. 6, pp. 161-180, 2001.

[22] R. Glass, I. Vessey und V. Ramesh, „Research in software engineering: an analysis of the literature," *Information ans Software Technology,* Bd. 44, pp. 491-506, 2002.

[23] V. B. Kampenes, T. Dybå, J. E. Hannay und D. I. K. Sjøberg, „A systematic review of quasi-experiments in software engineering," *Information and Software Technology,* Bd. 51, pp. 71-82, 2009.

[24] J. Siegmund, N. Siegmund und S. Apel, „Views on Internal and External Validity in Empirical Software Engineering," *Proceedings of the 37th IEEE/ACM International Conference on Software Engineering, ICSE 2015, Florence, Italy, May 16-24, 2015. IEEE 2015, ISBN 978-1-4799-1934-5,* pp. 9-19, 2015.

[25] P. Silapachote und A. Srisuphab, „Gaining and maintaining student attention through competitive activities in cooperative learning A well-received experience in an undergraduate introductory Artificial Intelligence course," *Proc. IEEE Global Engineering Education Conference,* pp. 295-298, 2014.

[26] S. Espana, N. Condori-Fernandez, A. Gonzalez und O. Pastor, „Evaluating the Completeness and Granularity of Functional Requirements Specifications: A Controlled Experiment," *Proc. of RE,* pp. 161-170, 2009.

[27] M. Genero, J. Cruz-Lemus, D. Caivano, S. Abrahao, E. Insfran und J. Carsi, „Assessing the Influence of Stereotypes on the Comprehension of UML Sequence Diagrams: A Controlled Experiment," *Proc. of MoDELS,* pp. 280-294, 2008.

[28] F. Ricca, M. Di Pinta, M. Torchiano, P. Tonella und M. Ceccato, „The Role of Experience and Ability in Comprehension tasks supported by UMLStereotypes," *Proc. of ICSE,* 2007.

[29] J. Carver, L. Jaccheri, S. Morasca und F. Shull, „A checklist for integrating student empirical studies with research and teaching goals," *J Empir. Sw Eng.,* pp. 35-59, April 2009.

[30] D. Port und D. Klappholz, „Empirical Research in the Software Engineering Classroom," *Proc. of 17th Conf. on Software Engineering Education and Training,* pp. 132-137, 2004.

[31] B. Boehm und S. Koolmanojwong, „Combining software engineering education and empirical research via instrumented real-client team project courses," *IEEE Conf. on Softw. Eng. Education and Training,* pp. 209-211, 2014.

[32] M. Höst, „Introducing Empirical Software Engineering Methods in Education," *Proc. of 15th Conf. on Softw. Eng. Education and Training,* pp. 170-179, 2002.

[33] M. Zelkowitz und D. Wallace, „Experimental Validation in Software Engineering," *Information and Software Technology 39(11),* November 1997.

[34] P. Berander, „Using Students as Subjects in Requirements Prioritization," *Proc. of ISESE,* 2004.

[35] F. Ricca, M. D. Penta, M. Torchiano, P. Tonella und M. Ceccato, „The Role of Experience and Ability in Comprehension Tasks supported by UML Stereotypes," *Proceedings of the 29th International Conference on Software Engineering (ICSE'07),* pp. 375-384, 2007.

[36] M. Daun, A. Salmon, T. Weyer und K. Pohl, „The impact of students' skills and experiences on empirical results: a controlled experiment with undergraduate and graduate students," *Proceedings of the 19th International Conference on Evaluation and Assessment in Software Engineering, EASE 2015, Nanjing, China, April 27-29, 2015,* pp. 29:1-29:6, 2015.